\documentclass[aip, jap, reprint]{revtex4-1}

\usepackage[colorlinks,bookmarksopen]{hyperref}
\usepackage{graphicx}
\usepackage{dsfont} 
\usepackage{xspace}
\usepackage{amsmath, amssymb}
\usepackage{epstopdf}
\usepackage{threeparttable}
\usepackage{tabularx}
\usepackage[english]{babel} 

\graphicspath{{/}}

\newcommand{\f}{\mathbf}                         
\newcommand{\txt}{\mathrm}                        
\newcommand{\ket}[1]{\left| #1 \right\rangle}     
\newcommand{\li}{\left\langle}
\newcommand{\re}{\right\rangle}
 
\newcommand{\matrixelem}[3]{\left\langle #1 \right| #2 \left| #3
\right\rangle} 
\newcommand{\ie}{\mbox{i.\,e.}\xspace}
\newcommand{\eg}{\mbox{e.\,g.}\xspace}
\newcommand{\eV}{\mbox{e\hspace{0.08mm}V}\xspace}          
\newcommand{\meV}{\mbox{me\hspace{0.08mm}V}\xspace}        

\newcommand{\imag}{\operatorname{Im}}

\begin{document}


\title{Tight-binding branch-point energies and band offsets for
 cubic InN, GaN, AlN and AlGaN alloys}



\author{Daniel Mourad}
\email{dmourad@itp.uni-bremen.de}
\affiliation{Institute for Theoretical Physics, University of Bremen,
Otto-Hahn-Allee 1, 28359 Bremen, Germany}


\date{\today}

\begin{abstract}
Starting with empirical tight-binding band structures,
the branch-point (BP) energies and resulting valence band offsets
(VBOs) for the zincblende phase of InN, GaN and AlN are calculated from 
 their $\f{k}$-averaged midgap
energy. Furthermore, the directional dependence of the BPs of GaN and AlN is
discussed using the Green's function method of Tersoff. We then show how
to obtain the BPs for binary semiconductor alloys within a band-diagonal
representation of the coherent potential approximation (CPA) and apply this
method to cubic AlGaN alloys. The resulting 
band offsets show good agreement to available experimental
and theoretical data from the literature.
Our results can be used to determine the band alignment in isovalent
heterostructures involving pure cubic III-nitrides or AlGaN alloys for arbitrary 
concentrations. 

\end{abstract}


\maketitle 


\section{Introduction\label{sec:introduction}}

Due to their wide range of potential application in optoelectronic  and
electronic devices, the III-nitride based semiconductor systems InN, GaN
and AlN and their ternary and quarternary alloys have attracted a great 
deal of interest for many years now.~\cite{ponce_nitride-based_1997}
They can crystallize in the thermodynamically stable
hexagonal configuration with a wurtzite crystal structure and in a
cubic modification with a zincblende
structure. Due to the low symmetry of the wurtzite system,
hexagonal nitrides do not only show strain-induced,
but also spontaneous polarization. The latter effect is absent
in the zincblende systems: it is for example possible to eliminate the
built-in fields in quantum dots by the growth of cubic III-nitride
along the nonpolar (001) direction, \eg by using GaN as dot and AlN as
barrier material.\cite{martinez-guerrero_self-assembled_2000,daudin_self-assembled_2001,gogneau_recent_2004,lagarde_exciton_2009,simon_direct_2003}
Furthermore, quantum well and superlattice structures based on cubic nitrides 
can be grown with good structural quality. AlGaN/GaN heterostructures can
\eg be used to create a two-dimensional electron gas without polarization
effects.\cite{wei_free_2012}

For the design as well as the theoretical simulation of such semiconductor
devices, the knowledge of the respective valence band offsets (VBOs)
across the junction of two semiconductor materials
A and B is of crucial interest.
In low-dimensional structures like superlattices, quantum wells or quantum dots,
the confinement potential is realized by 
a A/B heterojunction that results in a
suitable conduction band (CB) and valence band (VB) offset. Depending on the
values of the respective band gaps  and the resulting
VBO, the electrons and holes are either located in
the same (``type I'' alignment) or in different spatial regions (``type II''
alignment).

Furthermore, the VBO does not only determine the geometry
of the confinement potential and thus the electronic properties of
low-dimensional systems. It is 
also relevant for the simulation of alloyed systems of the type
A$_x$B$_{1-x}$, because the constituents have
to be aligned along a common energy reference scale to begin with. The 
concentration dependence of \eg the band gap will therefore directly 
depend on the energetic offset. In systems like the above mentioned
AlGaN/GaN, we even find a combination of these two effects.

Although many different experimental and 
theoretical methods are employed to determine VBOs,
the disparity in the literature values is often large.
On the theoretical side, computationally costful ab initio methods\cite{wei_calculated_1998,
moses_hybrid_2011} are ever-improving, but often still lack
quantitative agreement with the experiment and do not really 
provide theoretical insight into the underlying formation mechanism. 
A different approach is presented by models that predict
band discontinuities by alignment to a common energy reference like vacuum
levels,\cite{anderson_experiments_1962,van_vechten_quantum_1969,
van_de_walle_theoretical_1986,van_de_walle_band_1989}
or different kinds of charge neutrality levels.~\cite{harrison_tight-binding_1986,
cardona_acoustic_1987,van_de_walle_universal_2003,schleife_branch-point_2009,
hoffling_band_2010,monch_branch-point_2011}
This determination of transitive ``natural'' VBOs 
represents a rather macroscopical
approach that can give insight in the formation mechanism,
but neglects small-scale effects like lattice mismatch
or band bending on small length scales at the boundary. 

In the present work, we use two approaches from the literature
to determine the charge neutrality level, also known as branch point (BP), 
of the continuum
of interface states of cubic InN, GaN and AlN with band structures
from a flexible empirical tight-binding model (ETBM).
We then calculate the resulting VBOs between the III-nitrides and discuss
their directional dependence with the help of a Green's function method
from the literature.\cite{tersoff_theory_1984} The experimentally observed directional independence
of the cubic GaN/AlN VBO can nicely be explained in terms
of our results for the BPs.

As our results for the pure systems 
agree well with literature data (where available),
we use the coherent potential approximation (CPA) to calculate
the concentration-dependent BP position for disordered
cubic AlGaN alloys and compare
it to results from the simpler, but often applied
virtual crystal approximation (VCA). While the VCA fails to reproduce
the concentration dependence of the BP, the CPA results yield good agreement
with experimental data for Al$_{0.3}$Ga$_{0.7}$N/GaN interfaces. 
Our calculated CPA BPs can be used
to estimate the band alignment in interfaces with AlGaN alloys for
arbitrary concentrations.

\section{Theory\label{sec:theory}}

\subsection{Branch-point energy and band alignment
            \label{subsec:BranchPoints}}

The simplest linear models for the determination of VBOs across 
common interfaces between semiconductors use the alignment along a common 
reference level. An intuitive approach is the use of the vacuum level, 
\ie the comparison of the 
electron affinities of the constituent materials.\cite{anderson_experiments_1962}
However, the presence of interfaces and surfaces will influence the electronic 
properties, so that the use of values for infinite bulk crystals
 does not give satisfactory results for semiconductors. A large 
body of work deals with the refinement of this approach by establishing suitable
vacuum reference levels for model solids (see the excellent review part of
Ref.\,\onlinecite{lambrecht_interface-bond-polarity_1990} for further reference).

Another approach that renders good results is the use of a 
common charge neutrality level.
It is for example possible to align a broad range of materials along neutrality levels of
foreign atoms like interstitial hydrogen.~\cite{van_de_walle_universal_2003}
In a similar manner, it is well established to employ
the charge neutrality level of the continuum of interface
states. These localized states develop from itinerant Bloch states, but are characterized
by a  complex wave vector $\f{k}$, where $\imag(\f{k})$ is normal to the plane that
breaks the translational symmetry.
Consequently, they decay exponentially into the other
material and carry charge density across the interface,
which induces an interface dipole.~\cite{monch_electronic_2004}
The charge neutrality level of the spectrum of interface states is commonly
called the branch point  energy $E_\txt{BP}$
and is basically the crossover level between a donorlike and acceptorlike character.
The Fermi level $E_\text{F}$ will be pinned near the BP and the net charge density transfer
due to interface states will depend on the sign and magnitude of $E_\text{F} -E_\txt{BP}$.
The (hypothetical) exact alignment of the BPs of two materials 
with a common interface corresponds 
to a band lineup where all dipole contributions cancel each other out.

In Ref.\,\onlinecite{tersoff_theory_1984}, Tersoff  noted that $E_\txt{BP}$ can be 
calculated from the bulk properties of a semiconductor by the use of  the cell-averaged real-space
Green's function for the propagation along a lattice vector $\f{R}$.
This function 
is given as~\cite{allen_greens_1979} 
\begin{equation}
G^\f{R}_\txt{int}(E) 
    =  \sum\limits_{n \f{k}}
    \frac{e^{i \f{k}\cdot\f{R}}}{E-E_{n}(\f{k})},
\label{eq:greensfunction}
\end{equation}
with $E_n(\f{k})$ as the dispersion of the $n$-th band.
We denote this Green's function with the subscript ``int'' to
avoid confusion with the later introduced CPA Green's function,
although the definition (\ref{eq:greensfunction}) is per se independent 
of any interface. 

Because donor states predominantly take 
their spectral weight from the
VB, while acceptor states likewise originate from the CB,
$E_\txt{BP}$
can be identified as the energy where the CBs and VBs equally contribute to
$G^\f{R}_\txt{int}(E)$ when the interface normal
is chosen parallel to the direct lattice vector $\f{R}$.
As Tersoff pointed out, it is sufficient to  integrate over the
three-dimensional bulk Brillouin zone
(BZ) instead of separating $\mathbf{k}_\parallel$
and $\mathbf{k}_\perp$, as the choice of a large enough $\mathbf{R}$ automatically
projects out the relevant contribution to $G^\f{R}_\txt{int}(E) $.
This method is not 
applicable when the BP lies outside the band gap, however.

When the orientational dependence is small, the above mentioned approach
can substantially be simplified by the use of interface-averaged approximations.
In an attempt to generalize  previous 
approaches,~\cite{flores_energy_1979, tersoff_schottky_1985, cardona_acoustic_1987}
Schleife et
al. calculated the BP as a BZ average of the midgap
energy.~\cite{schleife_branch-point_2009}
 Reformulated in terms of the band-resolved densities of states (DOS)
 ~\cite{mourad_determination_2012}
$g_\txt{CB}^{i}$ ($g_\txt{VB}^{i}$) for a number of
 $N_\txt{CB}$ ($N_\txt{VB}$) CBs (VBs),
the BP is then approximated by  
\begin{equation}
\label{eq:dosbzaverage}
 E_{\txt{BP}} \approx
 \frac{1}{2} \int dE\,E \left[
    \frac{1}{N_\txt{CB}}  \sum\limits_i^{N_\txt{CB}}
  g_\txt{CB}^{i}(E)
  + \frac{1}{N_\txt{VB}} 
   \sum\limits_j^{N_\txt{VB}} g_\txt{VB}^j(E) \right].
\end{equation}
This formula also holds when the BP lies in a band, as experimentally known
for some  small-gap materials like InAs or InN.~\cite{noguchi_intrinsic_1991,
mahboob_origin_2004, piper_electron_2006} As no interface direction breaks
the symmetry in Eq.\,(\ref{eq:dosbzaverage}), the BP can in principle
also be obtained from the energies at the Baldereschi
point,\cite{baldereschi_mean-value_1973,chadi_special_1973} see \eg 
Ref.\,\onlinecite{monch_empirical_1996}.

It should be noted that both methods do not only require an appropriate
input band structure; in practice, the choice of a proper subset
of conduction and valence bands around the band gap region is also 
necessary.

Strictly speaking, the knowledge of
the energetic position of the BP is not sufficient to determine the VBO.
In real heterojunctions between two materials A and B,
the local charge neutrality condition will always be 
violated to some extent. The difference in the electron affinities results in 
a net dipole, which itself will be screened by a characteristic dielectric
response $\varepsilon$.
If the BP energy $E_\txt{BP}$ of each material is measured
with respect to its VB edge, the VBO $\Delta E_\txt{v} $
is obtained  as
\begin{equation}
\label{eq:lineupcondition}
\Delta E_\txt{v} = E_\txt{BP}^\txt{B} - E_\txt{BP}^\txt{A} + E_\txt{dip}.
\end{equation}
The magnitude of the interface dipole contribution $E_\txt{dip}$ can be estimated
following the work of M\"{o}nch.~\cite{monch_empirical_1996,monch_electronic_2004}
With $X_A$ and $X_B$ as electronegativity values from the Miedema scale, we have
\begin{equation}\label{eq:Miedemacorrection}
E_\txt{dip} = D_x(X_B - X_A),
\end{equation}
with the phenomenological slope parameter 
\begin{equation}
D_x = \frac{A_x}{1+0.1(\varepsilon_\infty-1)^2}.
\label{eq:slopeparameter}
\end{equation}

Here,  $A_x = 0.86$ \eV/Miedema-unit connects the work 
functions and the electronegativity scale, while  $\varepsilon_\infty$ is the
high-frequency dielectric constant of the semiconductor.
The Miedema electronegativity values were originally defined for elementary 
solids.~\cite{miedema_cohesion_1980}
Suitable values   for
binary compounds semiconductors can approximately be obtained from
the geometric mean of the constituents' values, \eg using  the data from
 Table A.4 of
Ref.\,\onlinecite{monch_semiconductor_2001}. Note that 
Eq.\,(\ref{eq:Miedemacorrection}) has originally been deduced for
metal-semiconductor contacts~\cite{monch_chemical_1996} but was then
found to be applicable to semiconductor
 interfaces, too.~\cite{monch_electric-dipole_2007}
 
\subsection{Empirical tight-binding model\label{subsec:ETBM}}

The importance of the choice of the band structure method in 
BP calculations has 
already been discussed in 
Ref.\,\onlinecite{mourad_determination_2012}. It has been pointed out 
that a suitably parametrized empirical tight-binding model will allow for a flexible
fit to a large parameter space; the input parameters can then be either
chosen from experiments or from results of more sophisticated
calculation schemes. They can as well deliver
a more or less realistic dispersion over the whole BZ, including
the Van Hove singularities at the zone boundaries, which give 
relevant contributions. 
Furthermore, due to the low computational cost, dense wave vector samples
samples on large $\f{k}$-space regions 
are effortlessly available,
 which is crucial to the convergence behaviour of  the Green function, 
Eq.\,(\ref{eq:greensfunction}).

Throughout this paper we will use a tight-binding model based on the 
work of Loehr.~\cite{loehr_improved_1994}
It makes use of a Wannier-like $sp^3$ basis per spin direction and lattice site
and can thus replicate eight doubly degenerate bands for crystals with zincblende structure.
The tight-binding matrix elements of the bulk Hamiltonian $H^{\text{bulk}}$  are
\begin{equation}\label{eq:EBOMme}
    E_{\alpha \alpha'}^{\mathbf{R} \mathbf{R'}} = 
                      \matrixelem{\f{R} \alpha}{H^{\text{bulk}}}{\f{R}'\alpha'},
\end{equation}
where  $\f{R}$ runs over the  sites of the underlying fcc Bravais lattice and
$\alpha$ is the orbital index.
The band structure $E_n(\mathbf{k})$ is obtained by  the diagonalization
of the matrix
\begin{equation}
    \sum_{\alpha'} \, \sum_{\mathbf{R}} e^{i \mathbf{k} \cdot \mathbf{R} }
    E_{\alpha  \alpha'}^{\mathbf{0} \mathbf{R}}
\label{eq:eigenproblem}
\end{equation}
for each $\f{k} \in $ 1.\,BZ.

By restriction of the  matrix elements to the second-nearest neighbor shell,
it is possible to fit the band structure to 11 material-specific parameters, which
are listed in Tab.\,\ref{tab:materialparameters} (the reader is referred to
Refs.\,\onlinecite{loehr_improved_1994, mourad_theory_2012,
 mourad_determination_2012} for a more
detailled discussion). The fact that all bands are also fitted to 
the $X$-point energies makes it especially
suitable for indirect materials, of which the properties are often not yet 
unambiguously
established. One notable example is AlN, which has a direct band gap
in its hexagonal modification,~\cite{vurgaftman_band_2003} while an indirect 
$\Gamma$-$X$ band gap is commonly~\cite{fritsch_band_2004}
 but not always\cite{fonoberov_excitonic_2003} assumed for the zincblende phase. 

Note that a modification of the model such as in   
Ref.\,\onlinecite{mourad_determination_2012} is not necessary here, 
as the spin-orbit interaction is comparatively weak in the III-nitrides,
which leads
to spin splittings two orders of magnitude smaller than the band gap.

In order to ensure the necessary consistency among the input values
(as \eg the energetic distance of bands influences their curvature due to
band-mixing effects) and 
simultaneously avoid the common band gap problem in ab initio calculations,
we use a comprehensive set of material parameters from
Fritsch et al.\,from empirical pseudopotential calculations
(Ref.\,\onlinecite{fritsch_band-structure_2003} 
for GaN and AlN, Ref.\,\onlinecite{fritsch_band_2004} for InN).
The input parameters for
the pseudopotentials stem from various sources which are listed
in the mentioned references. In case of InN, we additionally use recent
OEPx(cLDA) + G$_0$W$_0$ band curvature parameters
$m_c, \gamma_1, \gamma_2, \gamma_3$ from 
Rinke et al.,~\cite{rinke_consistent_2008} as their ab initio band
gap for zb-InN does only deviate slightly from experimental results.
Furthermore, we followed the advice of Vurgaftman and Meyer and corrected
the AlN gap slightly upwards,~\cite{vurgaftman_band_2003} as the original 
value from Thompson et al. has originally been determined for room
temperature,~\cite{thompson_deposition_2001} while the remainder of the input 
parameters is assumed to be valid for low temperatures.

\begin{figure*}
\includegraphics[width=\linewidth]{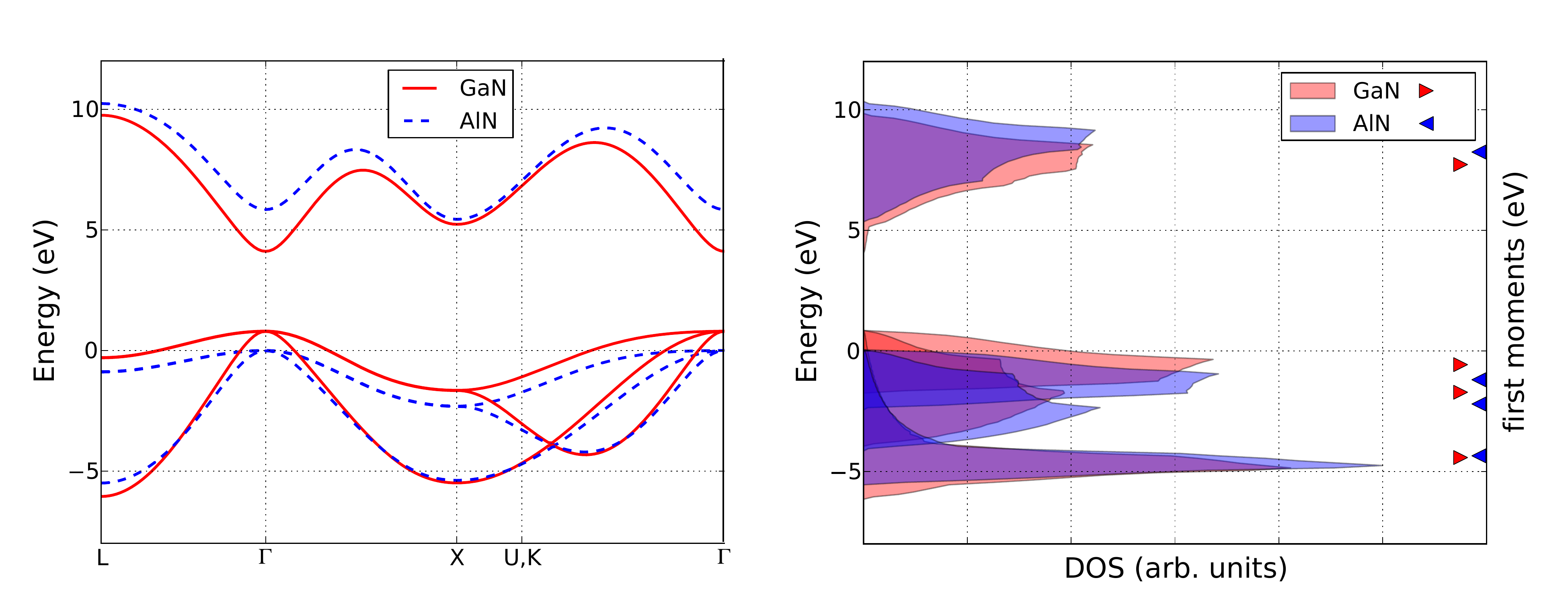}
\caption{(Color online) ETBM band structures (left) and band-resolved 
DOS and first moments (right) of zb-GaN and 
AlN. The later calculated VBO has already been anticipated here.}
\label{fig:alnganbsdos}
\end{figure*}

The resulting ETBM band structures and band-resolved densities of 
states for zb-GaN  and zb-AlN are depicted in Fig.\,\ref{fig:alnganbsdos}, together
with the first moments /centers of gravity of the bands
(the InN results are not added for the sake of clarity).

\begin{table*}
\centering
\caption{Input material parameters,
taken from Refs.\,\onlinecite{fritsch_band-structure_2003, fritsch_band_2004,
rinke_consistent_2008}. See text for further details.}
\label{tab:materialparameters}
\begin{tabular}{llrlll}
\hline
\hline
Parameter& Description &  & zb-AlN &  zb-GaN & zb-InN \\
\hline
$a$ & lattice constant & (\AA) &  \phantom{+}4.38  &
 \phantom{+}4.52 & \phantom{+}4.98\\
$\Delta_\txt{SO}$ & spin-orbit splitting & (\eV) &  \phantom{+}0.019
 & \phantom{+}0.017&  \phantom{+}0.006\\
$\gamma_1$ & Luttinger parameter &{}&  \phantom{+}1.85  & \phantom{+}2.89 & \phantom{+}6.82\\
$\gamma_2$ & Luttinger parameter &{}& \phantom{+}0.43  &  \phantom{+}0.85 &  \phantom{+}2.81\\
$\gamma_3$ & Luttinger parameter &{}& \phantom{+}0.74  &  \phantom{+}1.20&  \phantom{+}3.12\\
$m_\txt{c}$ & CB effective mass & ($m_0$) &  \phantom{+}0.23 & \phantom{+}0.14 & \phantom{+}0.05\\
$\Gamma_{1}^\txt{c}\,(\Gamma_{6}^\txt{c})$ & CB energy  & (\eV) &  \phantom{+}5.84 &  \phantom{+}3.31 &  \phantom{+}0.59\\
$\Gamma_{15}^\txt{v}$ & HH/LH VB energy & (\eV) & \phantom{+}0  &  \phantom{+}0 & \phantom{+}0\\
$X_1^\txt{c}\,(X_6^\txt{c})$ & CB energy & (\eV) & \phantom{+}5.44& \phantom{+}4.43 & \phantom{+}4.76\\
$X_5^\txt{v}$ & HH/LH VB energy& (\eV) & $-$2.32 & $-$2.46 & $-$1.48\\
$X_3^\txt{v}$ &split-off VB energy  & (\eV) & $-$5.39 &$-$6.29 & $-$4.80\\
\hline
\hline
\end{tabular}
\end{table*}

\subsection{Coherent potential approximation in band-diagonal representation 
            \label{subsec:CPA}}

The CPA\cite{soven_coherent-potential_1967, 
taylor_vibrational_1967,onodera_persistence_1968} is a 
well-established self-consistent
Green's function method  for the treatment of substitutionally
disordered systems.
The CPA itself and its broad
scope of application is excellently described 
in detail in Ref.\,\onlinecite{matsubara_structure_1982}.
For the sake of completeness, we will here only repeat the most important features.

In the standard form of the CPA, the TB Hamiltonian $H$ of the A$_x$B$_{1-x}$
alloy is seperated into a site-diagonal part $V$
and an off-diagonal part $W$, such that $H=V+W$. Here, 
only $V$ is 
considered to be site-dependent,
depending on the species on the site $\f{R}$.
The influence of the site-diagonal disorder is absorbed into 
the self-energy operator
\begin{equation}\label{eq:selfenergy}
\Sigma(z) \equiv H_\txt{eff}(z) - W,
\end{equation}
where $z$ is the complex energy and $H_\txt{eff}(z)$ is an effective
Hamiltonian; it is defined  such that the configurational average
$\li \ldots \re$ over the resolvent of $H$ equals the resolvent of
$H_\txt{eff}(z)$:
\begin{equation} \label{eq:Heff}
\li (z \mathds{1} - H)^{-1} \re \equiv \left[z \mathds{1} 
                                         - H_\txt{eff}(z)\right]^{-1}.
\end{equation}
When a representation is chosen,
this leads to a set of self-consistency conditions for the self energies, the
so-called CPA equations.
In the CPA, the self-energy operator is diagonal
in every representation. Additionally, its matrix elements
are neither dependent on $\f{k}$ nor on $\f{R}$.

The combination of the CPA with the here employed multiband 
ETBM has already been presented in 
Ref.\,\onlinecite{mourad_theory_2012}, together with a 
critical discussion of the extent of its validity. 
This reference also describes in detail how the CPA self-energy and Green's function
can  properly be transformed from the localized $sp^3$ basis to a band-diagonal
representation. The latter, in which $H_\txt{eff}$ is diagonal
in $\f{k}$, is crucial for the calculation of BPs for the alloyed
system, as one has to restrict the calculation to a subset of bands and the most
natural choice of band index $n$ is their energetic hierarchy.

In the here employed form, the CPA only maps the substitutional disorder in the 
first moments / centers of gravity $ E^\f{0}_\text{A/B}$ of the constituents on the 
self energy $\Sigma^n(z)$,
\begin{equation}
E^\f{0}_{n,\text{A/B}}   = \int dE E \, g^n_\text{A/B}(E)
                                 \stackrel{\text{CPA}}\longrightarrow \Sigma^{n}(z),
\label{eq:Wannierselfenergy}
\end{equation}
where $g^n_\text{A/B}$ is again the band-resolved DOS (normed to unity).
The off-diagonal disorder effects are simulated in the VCA, \ie interpolated linearly with the concentration $x$.
The self-consistent iteration scheme for the CPA equations as described in 
detail in Sec.\,II of Ref.\,\onlinecite{mourad_theory_2012} finally leaves us
with the CPA Green's function. Using its $\f{k}$-representation $G^{n\f{k}}_\txt{CPA}$, 
we can then extract the CPA one-particle spectral function $S^{n\f{k}}_{\txt{CPA}}$
and DOS $g^n_\txt{CPA}(E)$:
\begin{eqnarray}
S^{n\f{k}}_{\txt{CPA}}(E) &=& -\frac{1}{\pi} \lim\limits_{\delta \searrow 0}
            \imag G^{n\f{k}}_\txt{CPA}(E + i\delta), \label{eq:CPAspectralfunktion} \\
g^n_\txt{CPA}(E) &=&  -\frac{1}{\pi N_\f{k}} \lim\limits_{\delta  \searrow 0}\imag 
                   \sum\limits_\f{k}  G^{n\f{k}}_{\txt{CPA}}(E + i \delta).
                    \label{eq:CPADOS}
\end{eqnarray}
Here, $\delta \in \mathbb{R}$ and $N_\f{k}$ is the number of wave vector values
in the irreducible wedge.

Although the CPA can simulate many effects that go beyond the scope of
static mean-field treatments like the VCA (such as spectral broadening 
due to finite lifetimes or band gap
bowing), it also has several limitations.
Due to the translational invariance of $H_\txt{eff}$, the CPA cannot simulate 
local strain patterns. It will therefore only yield reliable results
if the difference
of the lattice constants is rather small. Besides this constraint, 
the CPA can be shown\cite{matsubara_structure_1982} to yield very good results 
in two limit cases:

(1) The \emph{weak scattering limit},
 where the difference of the 
first moments of the A and B bands is smaller than the respective bandwidths.

(2) The \emph{split band limit}, where these first moments are either far apart
and/or the bandwidths are small enough for the bands of A and B to not overlap.

\subsection{Calculation of branch points in the coherent potential approximation
            \label{subsec:CPABranchPoints}}

The approximated branch point energies for alloys can also be calculated from the
ETBM+CPA results.
We first define the complex quasi-particle band structure
\begin{eqnarray}
   E^{\text{CPA}}_n(\f{k}) &\equiv& \matrixelem{\f{k}n}{H_\txt{eff}}{\f{k}n}
                                                                  \nonumber\\
                           &=& \matrixelem{\f{k}n}{W}{\f{k}n}
                                + \Sigma^n(z = E^{\text{CPA}}_n),
                                                    \label{eq:quasiparticleBS}
\end{eqnarray}
where $\ket{\f{k}n}$ are Bloch states which diagonalize $H_\txt{eff}$.
Now the interface Green's function  
for the alloy can obviously be defined
by using $E^{\text{CPA}}_n(\f{k})$ as dispersion in
Eq.\,(\ref{eq:greensfunction}). 
As the CPA Green's function itself is given by
\begin{equation}
 G^{n\f{k}}_{\txt{CPA}}(z) =  \matrixelem{\f{k}n}{\left[z\mathds{1}
- H_\txt{eff}(z)\right]^{-1}}{ \f{k}n},%
\end{equation}
the interface Green's function can directly be obtained by means of 
a discrete Fourier transformation:
\begin{eqnarray}
G^\f{R}_\txt{int}(E) &=& \sum\limits_{n \f{k}} e^{i \f{k}\cdot\f{R}}
          {\left[E - E^{\text{CPA}}_n(\f{k})\right]^{-1}}\nonumber \\
          &=& \sum\limits_{n \f{k}}
               e^{i \f{k}\cdot\f{R}}\,G^{n\f{k}}_{\txt{CPA}}(E). 
          \label{eq:CPAinterfaceGF}       
\end{eqnarray}
Like in the pure case, the BP has to lie in a real gap of the quasi-particle
band structure. As $G^{n\f{k}}_{\txt{CPA}}$ has no poles in this 
region, the calculation can be restricted to the real part of the energy axis.

When the BZ average approach from Eq.\,(\ref{eq:dosbzaverage})
is used, the band-resolved densities of states can
analogously be replaced by their 
CPA counterparts from Eq.\,(\ref{eq:CPADOS}).

\section{Results\label{sec:results}}

\subsection{Pure zincblende III-nitrides
            \label{subsec:results_pure}}

\begin{table*}
\caption{Branch point energies $E_\txt{BP}$ for cubic III-nitrides in \eV.}
\label{tab:bpenergies}
	\begin{threeparttable}
	\begin{tabular}{cccccl}
	\hline
	\hline
	 Material &  $E_\text{g}$  &  \multicolumn{4}{c}{$E_\text{BP}$}  \\
	 \cline{3-6}
	 &          & $ \f{R} \propto (001) $ & $  \f{R} \propto (110)$ & BZ average &  Literature\\
	\hline
	zb-AlN & 5.44 & 3.38 & 3.64 & 3.28 & 2.97\tnote{a}, 3.19$\pm$0.18\tnote{b}\\
	zb-GaN & 3.31 & 2.62 & 2.86 & 2.50 &2.37\tnote{a}, 2.34$\pm$0.09\tnote{b}, 2.31\tnote{c}\\
	zb-InN & 0.59 &  --- & --- & 1.65 & 1.51\tnote{a}, 1.59$\pm$0.26\tnote{b}, 1.50\tnote{d}, 1.38$\pm$0.1\tnote{e}\\
	\hline
	\hline
	\end{tabular}
	\begin{tablenotes}
	\item[a]{ETBM + BZ average calculation, Ref.\,\onlinecite{monch_empirical_1996}, no explicit specification on bulk parameters.}
	\item[b]{Empirical value from fit to experimental band offsets, Ref.\,\onlinecite{monch_branch-point_2011}.}
	\item[c]{Quasiparticle BZ average calculation with $E_\text{g} = 3.28$ \eV (HSE03+$G_0W_0$), Ref.\,\onlinecite{schleife_branch-point_2009}.}
	\item[d]{Quasiparticle BZ average calculation with $E_\text{g} = 0.47$ \eV (HSE03+$G_0W_0$), Ref.\,\onlinecite{schleife_branch-point_2009}.}
	\item[e]{XPS measurement on (001) surface at room temperature,  Ref.\,\onlinecite{king_universality_2007}.}
	\end{tablenotes}
	\end{threeparttable}
\end{table*}

In order to calculate the BP energies from the TB band structures, 
we follow the recommendations of 
Schleife et al.\,~\cite{schleife_branch-point_2009} and use one
CB and the two uppermost VBs per spin direction. As  
 Fig.\,\ref{fig:alnganbsdos} suggests, this band subset can
be justified by the much larger dispersion of the lowest VB. It should 
however be kept in mind that this
set is obviously not suitable for several II-VI materials,
~\cite{mourad_determination_2012} and the
choice of bands certainly remains a major source of uncertainties in 
the BP calculation.
For that reason, 
we carefully recommend the same uncertainty range of 0.2~\eV as given
by Schleife et al. 

For the BZ average method, 
convergency of the results for $E_\txt{BP}$ is easily achieved
by using a moderately dense
$\f{k}$-grid on the irreducible wedge, so that a reduction to the
Baldereschi point does not render any practical advantage. Convergency within
10 \meV could be obtained with a equidistant resolution of
about 100--200 $\f{k}$-values
between $\Gamma$ and $X$.

\begin{figure}
\includegraphics[width=\linewidth]{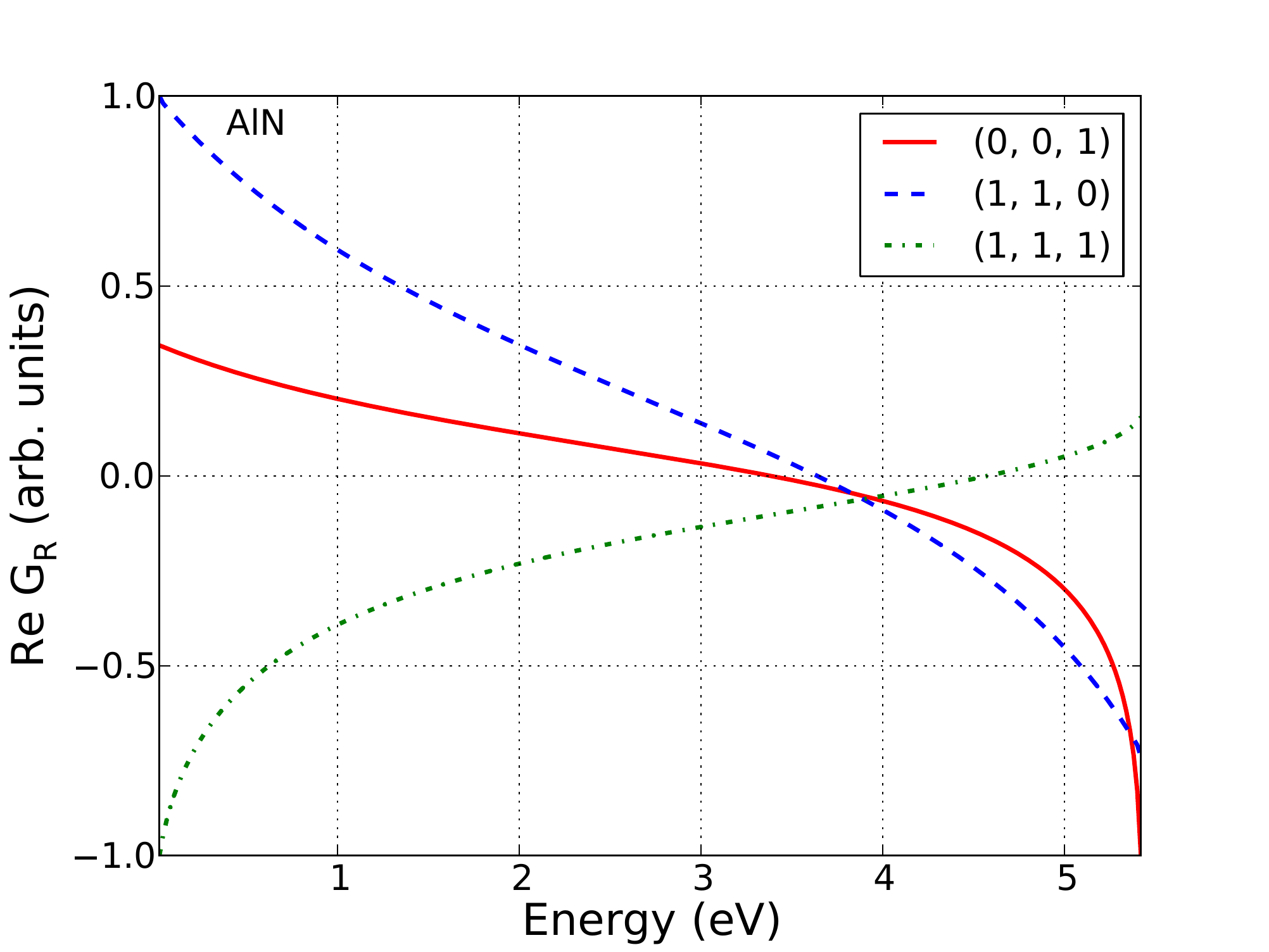}
\caption{(Color online) Real
part of $G^\f{R}_\txt{int}$ for AlN for
the (001), (110) and (111) direction.}
\label{fig:greensfunction}
\end{figure}

The interface Green's
function however is known to be extremely difficult to
converge.~\cite{schleife_branch-point_2009,mourad_determination_2012}
In practice, it turned out that very dense wave vector samples on
the full BZ are necessary. Figure \ref{fig:greensfunction}
exemplarily shows the real
part of $G^\f{R}_\txt{int}$ for AlN for three different 
interface orientations $ \f{R} \propto$ (001), (110) and (111).
As explained in Sec.\,\ref{subsec:BranchPoints}, 
the BP can be obtained by the zero-crossing
of this curve, which should arise for large enough multiples of 
$\f{R}$. For GaN and AlN, the intersections could be brought to convergency 
within 20 \meV for the (001) and (110)
direction when $\sim$250--300 $\f{k}$-values were
used between $\Gamma$ and $X$. However, satisfactory convergency for the (111)
direction could not be achieved for either material.

In case of InN, the results
from the BZ average method render a BP energy in the CB region,
so that the Green's function method is not applicable. This is also in agreement
with theoretical and experimental results from the
literature.~\cite{mahboob_origin_2004}

The results for the BP energies $E_\txt{BP}$ for the cubic III-nitrides InN, GaN
and AlN can be found in Tab.\,\ref{tab:bpenergies}, together with several
theoretical and experimental results from the literature. Further information
on the respective source is provided in the footnotes of the table. However,
it should be noted that the energy gap of the theoretical literature values
does in general not coincide with our input parameter set. Furthermore, no 
direction-dependent BP energies are available to our knowledge.
Still, our calculated values agree reasonably well with the external data.
For the sake of completeness, it should be noted that 
Ref.\,\onlinecite{belabbes_cubic_2011} contains further theoretical values for 
the cubic III-nitrides that were obtained from slightly different ab initio
band energies, but also show good agreement.

Our results also show that the position of $E_\txt{BP}$ carries
a clear-cut directional dependency, where 
\begin{displaymath}
E_\txt{BP}(\txt{BZ\,\,average}) < E_\txt{BP}(001) < E_\txt{BP}(110)
\end{displaymath}
for both
GaN and AlN. Additionally, the trend of the available data for the
(111) direction (see again Fig.\,\ref{fig:greensfunction}) also 
suggests that
\begin{displaymath}
E_\txt{BP}(111) > E_\txt{BP}(110) >E_\txt{BP}(001)
\end{displaymath}
might hold.
Whether the "real" directional average of the BP in these
materials lies energetically below the (001) and (110) values or
the BZ average method just gives a slightly too low $E_\txt{BP}$, 
cannot ultimately
be concluded from our calculations, as reliable results for further
directions would be necessary.

\begin{table*}
\caption{Resulting valence band offsets $\Delta E_\text{VB}(A/B)$
for cubic nitrides in \eV.}
\label{tab:purevbo}
    \begin{threeparttable}
    \begin{tabular}{cccccll}
    \hline
    \hline
     Interface A/B & &$E_\text{BP}^B - E_\text{BP}^A$ &  & Dipole &  \multicolumn{2}{c}{$\Delta E_\text{VB}(A/B)$} \\
     \cline{2-4} \cline{6-7}
    &  $ \mathbf{R} \propto (001) $ & $  \mathbf{R} \propto (110)$ & BZ average &  & This work & Literature \\
    \hline
    GaN/AlN & 0.76 & 0.78 & 0.78 & + 0.02 & 0.78, 0.80 & 0.5$\pm$0.1\tnote{a}, \,0.6\tnote{b}, \,0.8\tnote{c}, \,0.85\tnote{d} \\
    InN/AlN & --- & --- & 1.63 & + 0.03 & 1.66 & 1.1\tnote{c}, \,1.46\tnote{b}, \,1.60\tnote{d} \\
    InN/GaN & --- & --- & 0.85 & + 0.02 & 0.87 &  0.3\tnote{c}, \,0.75\tnote{d},  \,0.81\tnote{e}, \,0.86\tnote{b}\\
    \hline
    \hline
    \end{tabular}
    \begin{tablenotes}
    \item[a] Combination of intersubband and interband spectroscopies and ab initio HSE06 + $G_0W_0$ calculations on GaN/AlN quantum wells with lattice-matched AlN from Ref.\,\onlinecite{mietze_band_2011}.
    \item[b]{ETBM + BZ average calculation, Ref.\,\onlinecite{monch_empirical_1996}, no explicit specification on bulk parameters.}
    \item[c] Estimated after analysis of various experimental and theoretical sources,
    Ref.\,\onlinecite{vurgaftman_band_2003}.
    \item[d] Difference of branch points from Ref.\,\onlinecite{monch_branch-point_2011},
    see footnote of Tab.\,\ref{tab:bpenergies}. This should only be considered a rough 
    estimate, as the branch points were originally obtained the opposite way around.
    \item[e] Difference of branch points from Ref.\,\onlinecite{schleife_branch-point_2009},
    see footnotes of Tab.\,\ref{tab:bpenergies}.
    \end{tablenotes}
    \end{threeparttable}
\end{table*}

With knowledge of the BP energies, the band alignment can now be calculated
from Eqs.\,(\ref{eq:lineupcondition})-(\ref{eq:slopeparameter}).
For the contribution of the interface dipole, the
electronegativity values from Ref.\,\onlinecite{monch_semiconductor_2001} have 
been used. As the localized interface states exponentially decay into the energy
barrier caused by the VBO, we used the dielectric constant 
$\varepsilon_\infty(\txt{InN})=7.0$~\cite{schley_dielectric_2008}
for both the InN/AlN and InN/GaN junction
and $\varepsilon_\infty(\txt{GaN})=5.3$~\cite{levinshtein_properties_2001} for
GaN/AlN. The results for the BP difference, the dipole contribution and
the resulting VBO can be found in Tab.\,\ref{tab:purevbo}, together
with theoretical and experimental values from the literature.

For all interfaces, the contribution of the dipole correction is
small, so that an approximate alignment off these three systems along the
BP energy is well justified.
This is in agreement with general considerations in the literature
on junctions between cubic binary semiconductor 
compounds.~\cite{tersoff_theory_1984, monch_electronic_2004} 

For the GaN/AlN interface, the VBO is practically independent of the interface
orientation. This fact might seem strange at first glance, as the zincblende 
(001) interface is polar, while the (110) is not. However, this result
is in compliance with various experimental and theoretical findings 
(see \eg Ref.\,\onlinecite{vurgaftman_band_2003}) and has also been deduced
by Lambrecht and Segall~\cite{lambrecht_interface-bond-polarity_1990}
from an analysis of the bonding geometry and polarity at isovalent
zincblende interfaces. Our calculations show that this independence
can also be deduced from band-structure related properties, as the difference in
the (001) and (110) BP energies 
is almost identical (see again Tab.\ref{tab:bpenergies}).

Notably,
the BZ average method also gives the same BP difference and thus VBO,
which underlines its suitability for material systems with weak
directional dependence.
Although no
directionally dependent calculation could be performed for InN, the 
detailed analysis of Lambrecht and Segall indicates that the same behavior
will also hold for InN. 

From a quantitative point of view, the VBOs agree quite well with 
available data from the literature. Our value of 
$\Delta E_\txt{VB}(\txt{GaN/AlN}) \approx 0.8$\,\eV agrees  with the
recommendation from the topical review of Vurgaftman 
and Meyer~\cite{vurgaftman_band_2003} and also 
with the comprehensive experimental analysis of
M\"{o}nch.~\cite{monch_branch-point_2011} The theoretical
value of 0.6 \eV from M\"{o}nch also stems
from ETBM calculations, but with the use of atomic
term values and hopping matrix elements
from 1972 and 1981,
respectively.~\cite{fischer_average-energy--configuration_1972, harrison_new_1981}
The smallest
literature value of $0.5\pm0.1$ \eV has been determined from lattice-matched
quantum wells on 3C-SiC substrate at room 
temperature. For both InN/AlN
and InN/GaN, our results also agree well with the literature,
only the review by Vurgaftman et al.\,recommends smaller values.
This could be explained by the fact that their publication is 
from 2003, when high-quality wurtzite as
well as zincblende InN samples did not yet exist for a long time.

\begin{figure}
\includegraphics[width=\linewidth]{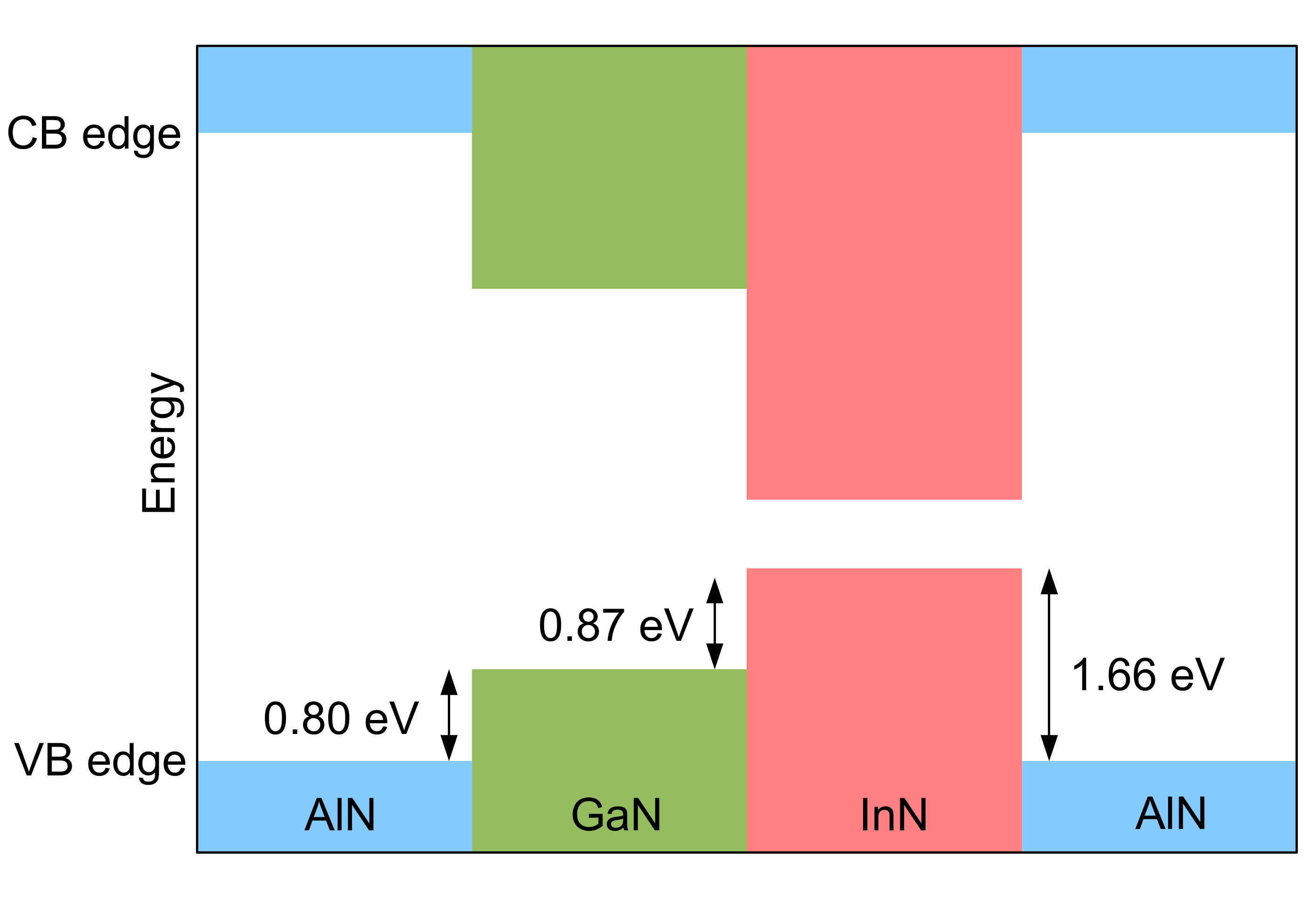}
\caption{(Color online) Band lineup for the zincblende modification
of the III-nitrides. The small deviation from the intransitivity 
between the GaN/AlN, InN/GaN and the 
InN/AlN VBO originates from interface dipole effects (see text).}
\label{fig:bandoffsets}
\end{figure}

The overall resulting band lineup as recommended by us for
the cubic phases of InN, GaN and AlN 
is depicted in 
Fig.\,\ref{fig:bandoffsets}.

\subsection{AlGaN alloys with zincblende phase\label{subsec:results_alloys}}

In contrast to InGaN and InAlN, the lattice mismatch between the
constituents in cubic AlGaN is relatively small with 3\%, which suggests
that the influence of local lattice distortions is not of major importance
when the electronic structure of the alloy is calculated.

In fact, it has
already been shown in Ref.\,\onlinecite{mourad_theory_2012} that the 
CPA gives good agreement with supercell calculations and further literature
values for the $\Gamma$-valley bowing and the crossover concentration
between a direct GaN-dominated and indirect AlN-dominated behavior. Besides
a small variation in the band curvatures, which is not relevant
for the here discussed properties, those 
calculations were performed for an almost identical parameter set
and the same VBO (albeit by chance, as the value of
$\Delta E_\txt{VB}(\txt{GaN/AlN}) \approx 0.8$\,\eV calculated in this paper
coincides with the there employed recommendation from Vurgaftman and Meyer).
A closer look at the bandwidths and the position of the first band moments
as depicted on the right of Fig.\,\ref{fig:alnganbsdos} shows that this
system is very close to the weak scattering limit, which also underlines
the applicability of the CPA. 
Furthermore, recent experimental data for the band alignment is available,
in particular 
for cubic nonpolar Al$_{0.3}$Ga$_{0.7}$N/GaN heterostructures, which were
grown by means of molecular beam epitaxy on 3C-SiC (001).
In Ref.\,\onlinecite{wei_free_2012}, Wei et al. report on a 
two-dimensional electron gas at this interface and observe
a large conduction-to-valence band offset ratio of 5:1 which 
enhances the electron accumulation. 

As the electronic structures of two materials A and B enter the calculation
for alloys, an energy offset and thus VBO between A and B has also
to be established as input parameter.
In general, the choice of a VBO for CPA or supercell calculations for 
alloys is far from trivial. With varying concentration $x$ in
the A$_x$B$_{1-x}$ alloy, the outline of the problem
does in principle change from single-impurity-like behavior over dilute 
systems up to whole A-in-B-clusters, so that the "true" VBO should in
principle be concentration- as well as lattice-site-dependent. Nevertheless,
the usage of a fixed parameter is common for the sake of simplicity. 
In case
of the AlGaN system under consideration, we will use the interface VBO of
 0.8\,\eV, as it has been shown
to be directionally independent in the last section and gives 
the correct limit in the pure and in the phase separation case.

\begin{figure}
\includegraphics[width=\linewidth]{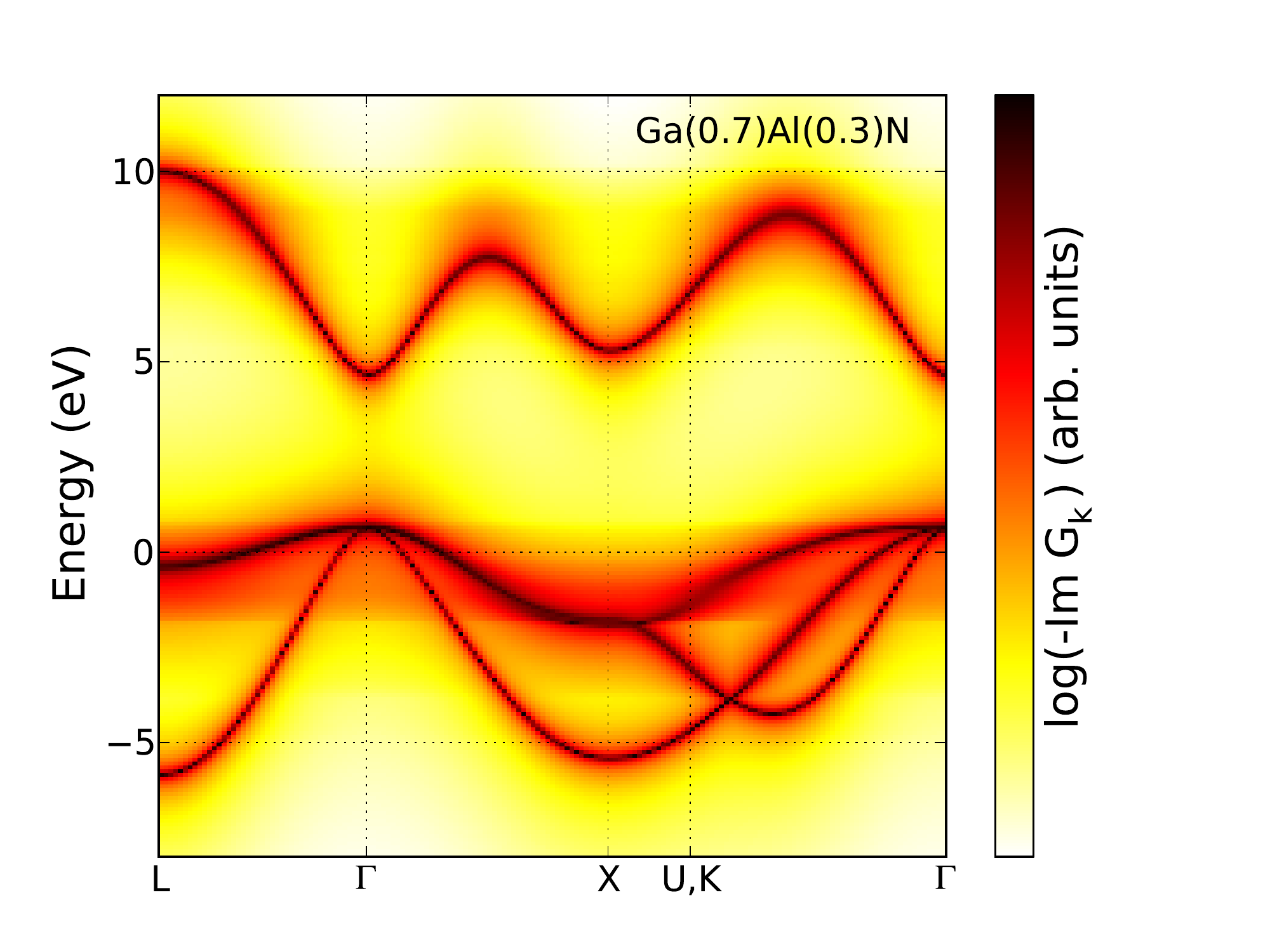}
\caption{(Color online) CPA quasiparticle band structure for
Al$_{0.3}$Ga$_{0.7}$N.}
\label{fig:algancomplexbs}
\end{figure}

The resulting CPA quasiparticle band structure for Al$_{0.3}$Ga$_{0.7}$N 
is depicted
in Fig.\,\ref{fig:algancomplexbs}. Note that the linewidths were artificially
enlarged for 
enhanced contrast in this figure.
For vanishing imaginary part in the complex energy
 $z=E + i \delta$, an unambiguous CB edge and VB edge and thus band gap 
can be identified.


In order to obtain the position of the BP over the whole concentration range,
a small imaginary part of 
$\delta=10^{-4}$\,\eV and $\#\f{k}\approx 10^7$ values in the irreducible
wedge have been used in the CPA iterations, together with an energy
resolution of $\Delta z = 5 \times 10^{-3}$ \eV.
In the following, we restrict the discussion to the results of the BZ
average approach as defined in Eq.\,(\ref{eq:CPADOS}) for two reasons:

(1) The almost identical energy difference for pure GaN and AlN between
$E_\txt{BP}(\txt{BZ\,\,average}), E_\txt{BP}(001)$ and $E_\txt{BP}(110)$
of 0.1 and 0.25 \eV, respectively,
directly carries over to the results for AlGaN when the quasiparticle
band structure 
is used according to Eq.\,(\ref{eq:CPAinterfaceGF}).
This holds in good approximation over the whole concentration range. Due to
the very slow convergency, this has only been spot-checked with a lower energy
resolution.

(2) In real materials, disorder effects will prevent any translational
invariance, which is an artificial symmetry of the CPA Hamiltonian.
It is therefore at least questionable whether effects that strongly
originate from 
symmetry properties of the quasiparticle band structure carry
over to the behavior
at real disordered interfaces. 

In addition, virtual crystal calculations
have also been performed for comparison,
as the VCA is also used in the literature
to calculate branch-point energies of alloys,~\cite{monch_empirical_1996}
supposedly due to its simplicity. Here, the matrix elements of the alloy
with concentration $x$
are obtained by a linear interpolation between the pure values. As the VCA
is contained in the CPA as a limit case,~\cite{matsubara_structure_1982}
the CPA results serve as a benchmark for the reliability of the VCA.

\begin{figure}
\includegraphics[width=\linewidth]{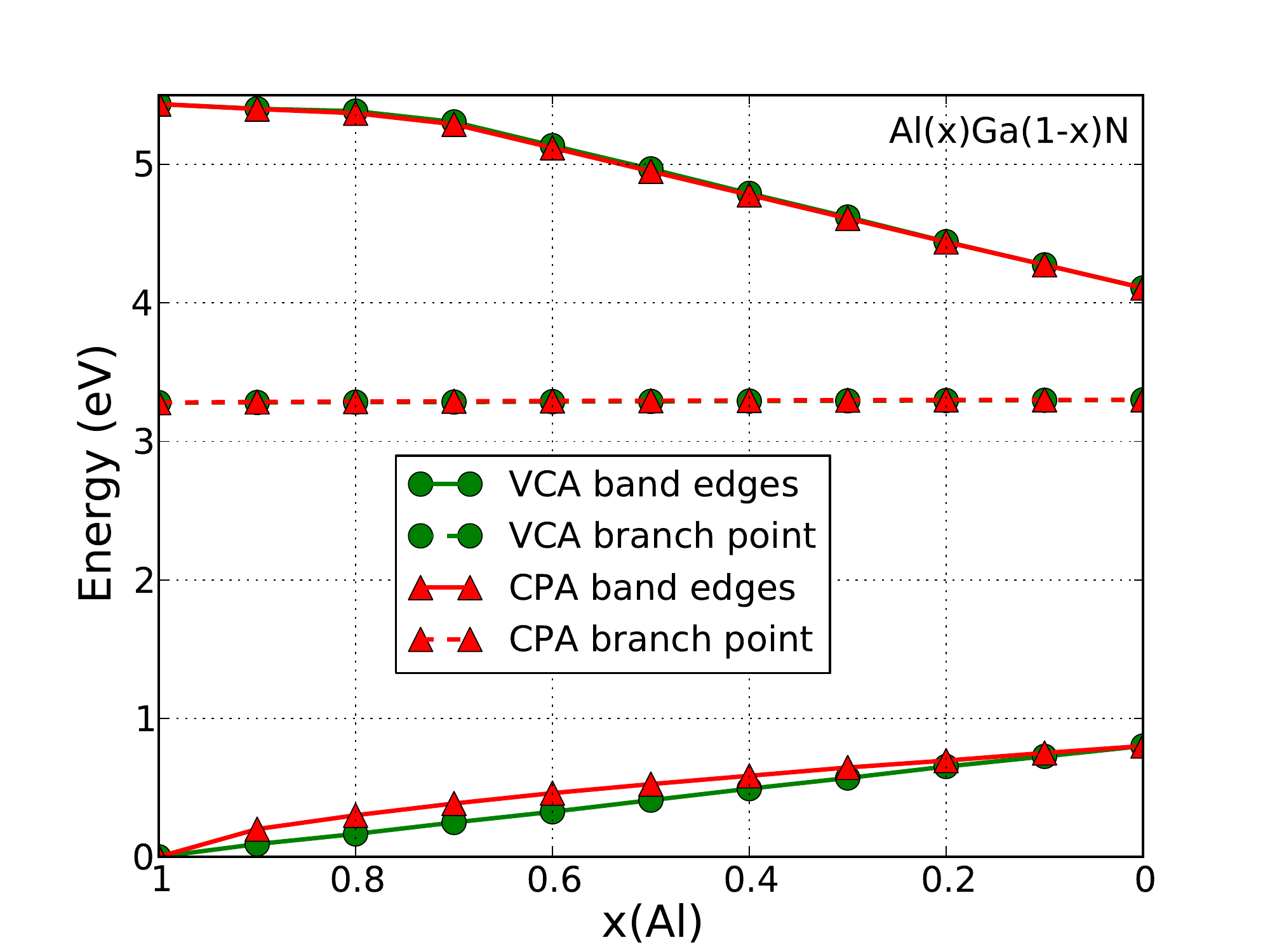}
\caption{(Color online). CB edge, VB edge and BP energy of AlGaN, calculated
in the CPA and the VCA from the TB band structures, respectively. 
The VCA results for the BP and the CB edge are overlaid by the CPA data.}
\label{fig:alloyedges}
\end{figure}

The resulting band edges and BPs of AlGaN can be found in
Fig.\,\ref{fig:alloyedges}. Here, the same band subset as in the
pure case was used for all concentrations,
but for material combinations close to the 
split-band limit, an individual and
concentration-dependent choice might be necessary.
It is immediately evident that the position of the CPA as well as the
VCA BP is almost constant on the common energy scale of the 
A$_x$B$_{1-x}$ alloy, so that the relevant 
value for band alignment, the energetic distance between the BP and the VB
edge, is only determined by the latter. This behavior originates in the use
of a constant A-to-B VBO that has itself been determined by the alignment
of BPs and will not hold when either a concentration-dependent
VBO and/or a VBO with large dipole contribution is used in the calculations.

\begin{figure}
\includegraphics[width=\linewidth]{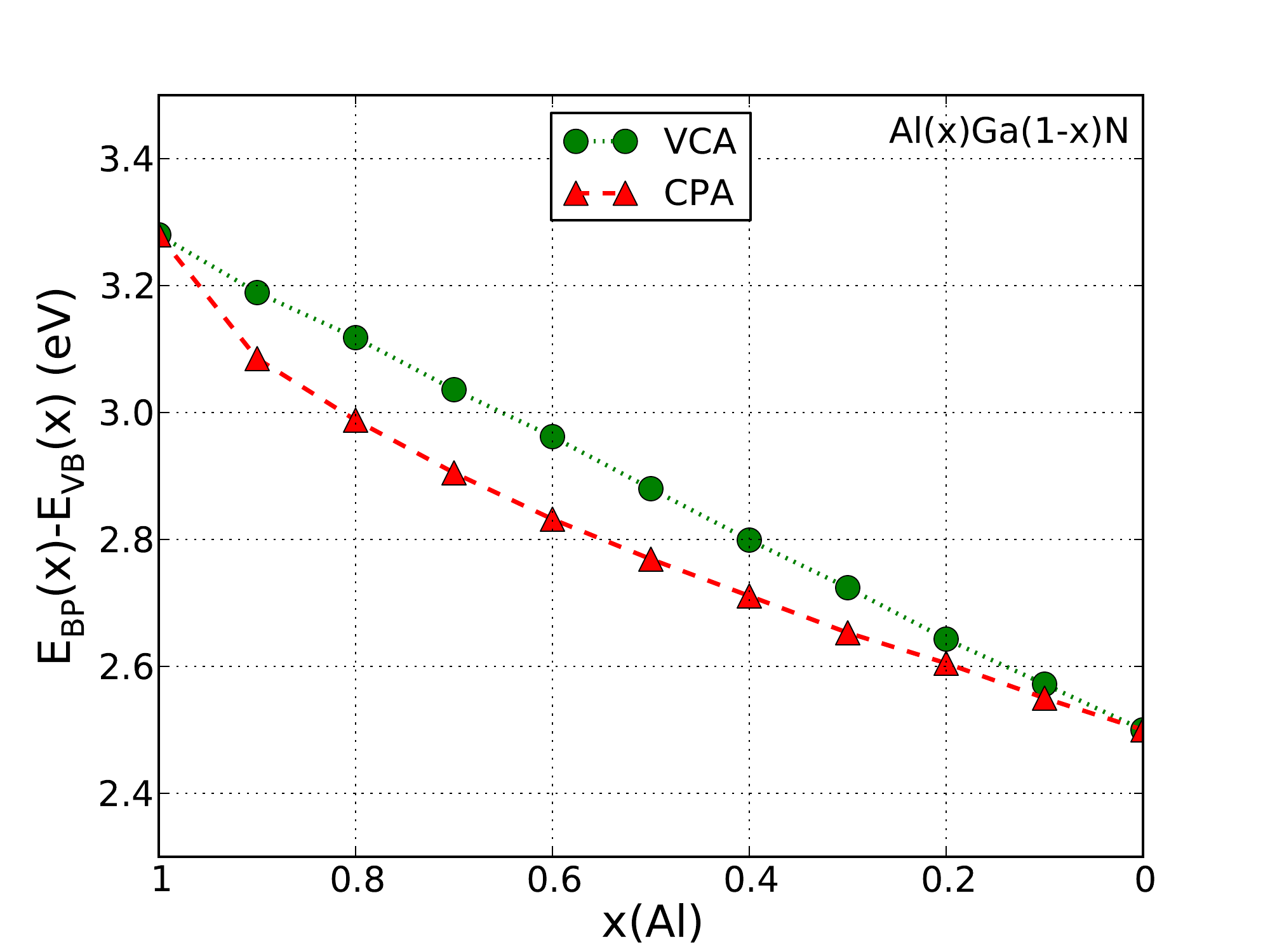}
\caption{(Color online) Relative BP position
$E_\txt{BP}(x)-E_\txt{VB}(x)$ of AlGaN calculated in the CPA and the VCA.}
\label{fig:bpbowing}
\end{figure}

A comparison of the CPA to the VCA results reveals that the latter
gives slightly too low VB edges.
In Fig.\,\ref{fig:bpbowing}, the relative BP position
$E_\txt{BP}(x)-E_\txt{VB}(x)$ is
depicted as a function of the concentration $x$. While the VCA
erroneously yields an almost linear behavior, the relative CPA BP position shows
a pronounced bowing, where a satisfactory fit by a quadratic or cubic function
is not possible for the whole concentration range. 
Table \ref{tab:energeticpositions} comprehensively lists the band edges
and the relative BP energy of zb-AlGaN.
These values can be used to estimate the band alignment for all
concentrations with respect to 
any material whose branch points are known. Like in the pure case,
the uncertainty will at least be 0.2 \eV, not including additional
deviations due to the use of the CPA.
 
\begin{table}
\centering
\caption{Concentration-dependent VB and CB edge $E_\text{VB/CB}$,
and relative BP position
$E_\text{BP}-E_\text{VB}$ for zb-AlGaN as obtained in the CPA.
All values in \eV.}
\label{tab:energeticpositions}
  \begin{threeparttable}
  \begin{tabular}{llllc}
  \hline
  \hline 
  & $x\,(\text{Al})$ & $E_\text{VB}$ & $E_\text{CB}$ & $E_\text{BP}-E_\text{VB}$ \\
  \hline
  AlN & 1   & 0 & 5.44\tnote{$\dagger$} & 3.28 \\
      & 0.9 & 0.20 & 5.40  & 3.09 \\
      & 0.8 & 0.30 & 5.37  & 2.99 \\
      & 0.7 & 0.39 & 5.29  & 2.91 \\
      & 0.6 & 0.46 & 5.12  & 2.83 \\
      & 0.5 & 0.53 & 4.95  & 2.77 \\
      & 0.4 & 0.59 & 4.78  & 2.71 \\
      & 0.3 & 0.65 & 4.61  & 2.65 \\
      & 0.2 & 0.70 & 4.44  & 2.61 \\
      & 0.1 & 0.75 & 4.28  & 2.55 \\
  GaN & 0   & 0.80 & 4.11\tnote{$\dagger$} & 2.50 \\
  \hline
  \hline
  \end{tabular}
  \begin{tablenotes}
  \item[$\dagger$]Input values from parametrization, see Tab.\,\ref{tab:materialparameters}.
  \end{tablenotes}
  \end{threeparttable}
\end{table}

We will finally compare our results to the work of Wei et al. for cubic
Al$_{0.3}$Ga$_{0.7}$N/GaN heterostructures.~\cite{wei_free_2012} 
The difference
in the electrostatic potential at the interface, which is determined by means of
electron holography, gives a CB offset of 0.65 \eV in their experiment.
Together with a measured band gap difference of 0.78 \eV across the junction, this
yields a VB offset of $0.13$ \eV. The value is in
very good agreement with our theoretical result of
$2.65\,\eV-2.50\,\eV=0.15\,\eV$, However, this compliance definitely
exceeds the predictive
power of the calculations.
When combined, our
BP and band gap results for this material combination give a 
conduction-to-valence-band ratio of $0.5/0.15 \approx 3.3$, which is 
below the experimental value of 5.
The deviation can be traced back to their spectroscopically determined
band gap difference between Al$_{0.3}$Ga$_{0.7}$N and GaN, which is larger
than our calculated band gap difference of 0.65 \eV. Also,
our calculations do not include any effects from lattice strain, which,
 albeit small, will also influence the results.
Notably, the AlGaN layer thickness in their experiment is only 30 nm,
so that finite 
size effects will also likely play a role. Furthermore, the authors notice
composition fluctuations and oxygen contamination in their samples.  
If we account for these effects by using the same modified 
band gap difference of 0.78 \eV in our calculations,
we arrive at a ratio of 4.2, which is surprisingly good, 
especially
when the various sources of uncertainty are considered on both the
theoretical and the experimental side.

\section{Summary\label{sec:summary}}

In this paper, we showed that the branch-point (BP) energies and resulting
band alignment for the cubic III-nitrides InN, GaN and AlN 
can be calculated from tight-binding band structures with consistent
and up-to-date input parameters. The directional
independence of the valence band offset (VBO)
 at the unstrained GaN/AlN
interface has been traced  back to a constant shift 
of the corresponding BPs by the use of Tersoff's Green's function 
method. We then showed how the coherent potential approximation (CPA) 
can be used in combination
with the tight-binding model to obtain the BPs and VBOs for alloyed systems and 
applied this method to the zincblende modification of AlGaN over the 
whole concentration range. While
a virtual crystal treatment of the system gives erroneous results,
the CPA agrees well with experimental data
on Al$_{0.3}$Ga$_{0.7}$N/GaN heterostructures. 
Our results
can be used to determine the band alignment in isovalent heterostructures
involving pure cubic III-nitrides or AlGaN alloys for arbitrary concentrations.

\begin{acknowledgements}
The author would like to thank G.\,Czycholl for interesting discussions and
the introduction to the CPA.
The author also thankfully acknowledges the developers of the free Python
numpy and matplotlib modules, which were used for calculation and visualization.
\end{acknowledgements}



%

\end{document}